\documentstyle[epsfig]{aipproc}

\begin{document}
\title{Gravitational clustering\\in N-body simulations }

\author{Maurizio Bottaccio$^1$, Roberto Capuzzo Dolcetta $^1$, Paolo Miocchi$^1$,
Alessandro Amici$^1$, Marco Montuori$^2$ and Luciano Pietronero$^{1,2}$}
\address{$^1$ Physics Department, University of Roma "La Sapienza", Roma, Italy\\
$^2$ INFM Section of Roma 1, Roma, Italy}

\maketitle

\begin{abstract}
In this talk we discuss some of the main theoretical problems
in the  understanding of the statistical properties of gravity.
By means of N-body simulations we approach the problem of understanding 
the r\^ole of gravity in the clustering of a finite set of 
N-interacting particles which samples a 
portion of an infinite system. 
Through the use of the 
conditional average density,
we study the evolution of the clustering for the system
putting in evidence some interesting and not yet 
understood features of the process.

\end{abstract}

\section*{Introduction}
 
The study of the features of gravitational clustering is one 
of the most challenging problems of classical physics, and an interesting
example of interplay between statistical mechanics and astrophysics. 

From the point of view of statistical mechanics, it is very
hard to study the properties of  an infinite system of self-gravitating 
particles. This is mainly due to the long range
nature of gravitational potential, which is not shielded
by the balance of far away charges, as e.g. in a  
plasma. Therefore all scales contribute to the evaluation of
the potential energy of a particle. 
As a consequence of such difficulties, 
a satisfying thermodinamic equilibrium treatment
of such systems is still lacking. 

Most studies on self gravitating systems have been focusing on clustering
properties of well-defined physical systems: from interstellar 
cold molecular clouds 
 to globular clusters, to clusters and superclusters of galaxies.
The range of scales on which such processes take place is impressive 
(from $10^{-1} pc$
to $10^8 pc$). This also implies interactions with physical processes of different
types, depending on the scale (from e.g. turbulence in cold molecular clouds 
to cosmological expansion above galaxy scale). 

In spite of the large amount of analytical and numerical studies on 
gravitational clustering, it is very difficult to retrieve a clear picture of
the statistical properties of self-gravitating systems, 
exactly because of the richness of 
physical processes which they take into account.

What has especially raised our interest in the subject was the results
of  a series of analyses of space correlations in 
galaxy catalogues by Pietronero et al. \cite{Pietro}. 
They revealed that galaxies form a fractal set with well defined properties
at least up to a certain scale.
The discovery of the fractal characteristics of large scale structures in the 
universe raises a number of new and extremely appealing questions, which have
neverthless been almost completely neglected.
For example we can ask  whether gravity is a self organizing process, or if
the formation of fractal properties depend on the choice of  a particular set
 of initial conditions, or
even one has to invoke some other physical process.

In a more general perspective, we intend to study 
in detail the characteristics
of the evolution of self gravitating systems, {\em from the point of view
of statistical mechanics}.
In particular we focused on  the 
evolution of spatial correlations in self-gravitating systems, and on the
possibility that they may develop some kind of self similar spatial 
or temporal feature.  

N-body simulations represent in fact a valuable experimental setting for a 
careful investigation of the dynamics of clustering. 
Current astrophysical simulations have reached a high level of refinement, both in resolution 
and in the number of different physical processes which they take into account.
Such characteristics allow them to study in great detail the single physical 
problem for which they are developed. 
On the other hand they don't allow a 
clarification the common r\^ole and the peculiarities of gravitational 
interaction. 

Our approach has been {\em completely different}: we start from the
simplest possible model and study in detail what happens during the clustering. 
Even {\em such a simple model} gives raise to a rich phenomenology, 
which has
{\em not been studied in a systematic way} up to now. 

\section*{Theoretical issues on gravitational clustering}

 The peculiar  form of the gravitational potential makes the 
statistical properties of self-gravitating systems a very difficult  subject to
study.
Two classes of problems arise: 
 those due to 
the {\em short range} (i.e. $r\rightarrow0$)
divergence 
and those due to the {\em long range} 
(i.e. $r\rightarrow +{\infty}$) behaviour.
 
The former is not uncommon, since it  is the same problem
which arises in electromagnetism. The divergence would cause, e.g.,
the Boltzmann factor to diverge in the limit $r\rightarrow0$.
A typical prescription is to put a small distance 
cut-off in the potential. The physical nature of this cut-off may be 
due to many effects, e.g.  the dynamical
emergence of angular momentum barriers. 

The long range behaviour is of much more concern and is, in fact,
{\em the} problem. 
It is an easy exercise to verify that the energy
of a particle in an infinite self-gravitating system diverges. 
This causes the energy to be {\em non-extensive}.
As a consequence,  a thermodynamical limit is not achieved,
since as the number of particles goes to infinity while keeping 
the density constant, the energy {\em per particle} diverges.
Strangely enough, such a problem has not been fully appreciated
by  many physicists in the field (see e.g. \cite{Padma}), as they
try to avoid the long range divergence by putting the system in 
a box ``as it is usually done with ordinary gas''. In fact, the difference is
that in ordinary gas, when confining the system in a box, the energy
per particle is equal to a constant plus a surface term that goes
to zero in the thermodynamical limit. In self-gravitating systems,
due to non extensivity, the energy per particle is neither a constant,
nor the surface term goes to zero (in fact, it is of the same order of 
magnitude as the potential energy due to particles belonging to the system). 

Another very interesting consequence, which is often not appreciated,
is that the thermodynamical definition of temperature, as the parameter
which controls the equilibrium of the system, doesn't hold for a self 
gravitating system, since one cannot divide a system into smaller
subsystems with the same thermodynamic properties of the larger system. 

So far we have discussed the problems one encounters when one 
tries to build a theory for thermodynamic equilibrium. 
However we are much more interested in what happens {\em out of equilibrium},
during the evolution of a system. For example it is highly probable
that {\em any fractal state could not survive} when  the system would 
reach the
equilibrium.
Problems due to long range interactions are present all the same
in most approaches,
but they are more subtle to be put in evidence.

\section*{N-body simulations}

N-body simulation is one of the main tools for the study of gravitational
clustering. They provide in fact the possibility to test theoretical 
models and to study the evolution of selfgravitating systems, a feature
that is very hard to obtain from observations.
Several different algorithms are used (see \cite{nbdrev} for a review).

A brute force method to evaluate forces acting on particles in a simulation
requires $O(N^2)$ computations, a feature which forbids them to be useful
if large number of particles are required.
The tree algorithm (see \cite{barnes}, and \cite{alltree} for a detailed 
explanation) is an 
approximation scheme which involves $O(N log N)$ computations, and whose 
level of accuracy can be tuned to match the problem requirements.

It is based on the decomposition of the simulation volume into a tree-like
 structure. Particles are grouped with well-defined rules 
in a hierarchy of clusters. 
The contribution of 
distant clusters to the force on a particle is taken into account 
by a multipole expansion of the
potential generated by them. This approximation greatly reduces the 
computational expense of a simulation. 

The algorithm has to be elaborated for our purposes by adding periodic
boundary
conditions, which allow a statistically
equivalent environment for all the particles, and also  include the 
long range interactions.

In our simulation we started from the simplest possible model:

i) random (white noise) initial positions of particles;

ii) no cosmological expansion;

iii) zero initial velocities;

iv) equal mass particles.

The main tool we use to analyse spatial correlation of the system 
is to measure the {\em conditional average density} $\Gamma^*(r)$. 
It is defined as 
$\Gamma^*(r)=<N(r)>/V(r)$ 
where $<N(r)>$ is the average number of points (not considering the center) 
in a sphere of radius $r$ and volume $V(r)$
centered on each point of the system.

\begin{figure}[b]
\vspace{30pt}
\centerline{\epsfig{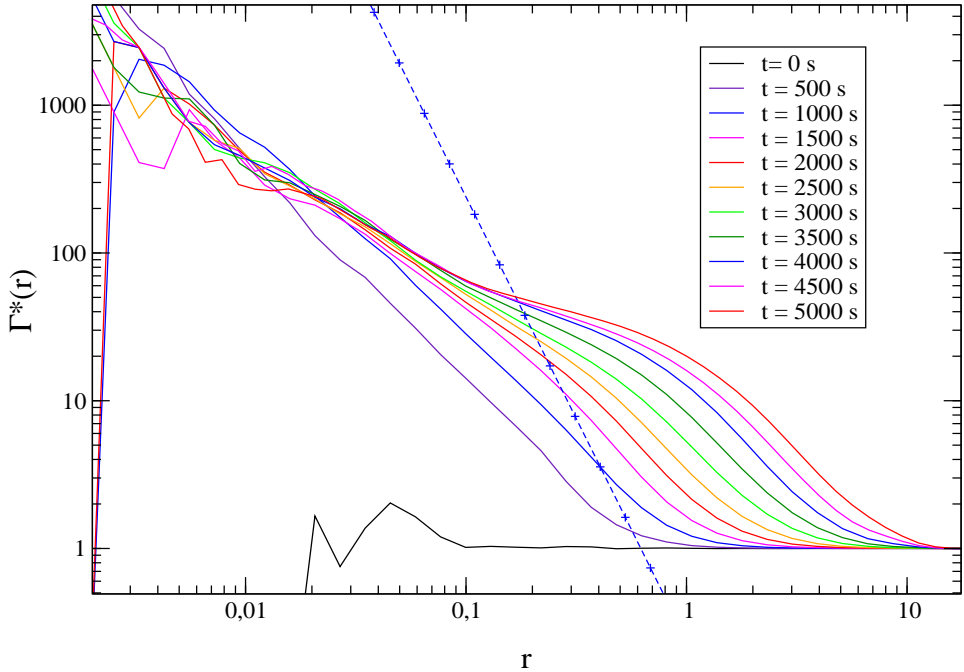}}
\caption{Evolution of $\Gamma^*(r)$ in a simulation with $32^3$ 
particles. Measures are taken each $500 s$. $\Gamma^*(r)$ 
moves from left to right with time. Left of the dash-crossed 
line we have less than one particle on average in a sphere of radius 
$r$ centered on a particle (the particle at the centre is not included).}
\label{fig1}
\end{figure}

\section*{Results}
All the simulations we have performed have the same average number density,
$\bar n =1$ which sets the physical scale of the system.

Figure \ref{fig1} shows the evolution of $\Gamma^*(r)$ with time for a 
simulation of $32^3$ particles. 
This picture shows some interesting features, and allow us to focus on the 
different aspects of gravitational clustering. We can identify 
the region left of the dash-crossed  line as a regime of 
mainly two 
particles interactions. In fact for such small scales, we find on average less than one
particle in each sphere centered on each particle of the system.
This region shows a very nice power law 
behaviour which seems stable for quite a long time,
Such behaviour is not reported in the literature, 
as far as we know. 
The {\em growing front} of the correlations shows beautiful features 
of ``temporal self-similar behaviour'', in the sense that if  
we rescale length scales
with a function of time the curves collapse.
The scaling factor is exponential. 
It is interesting to notice that while
the exponential time dependence can be naively 
retrieved (with the right coefficients)
 using the solutions for the linear theory of the growth 
 of density perturbation in a self gravitating fluid, the behaviour of
the correlations is not well described by such solutions.
In fact they  predict 
a selfsimilar (in time) growth of the initial correlations of
density fluctuations, 
until $\delta n/ n< 1$. On the other hand, an inspection of the fig.\ref{fig1} 
shows that such growth takes place at  large number density contrasts (i.e. $\delta n/ n>1$),
and does not correspond to a pure amplification of the  initial conditions. 
Anyway, such theory is considered very well verified
in cosmological simulations. 
There may be several explanations. One possibility is that some
of the approximations used to derive such solutions break earlier without 
expansion. Another interesting remark could be that in our simulations
we don't suppress the {\em granular} features of particles with large smoothing
length, as it  is customary in many cosmological simulations
(see e.g. \cite{Theis}). 
Such an observation paves the way to a very interesting
issue which we are currently investigating: 
whether the  granularity is an essential 
feature  to the understanding of  gravitational clustering, 
or which is the limit of 
validity of the fluid description of a self gravitating medium.
Another region of interest would be the one between the two we have just 
described. 
With present data, we cannot resolve the behaviour of correlations in 
such regime, due to  small numbers of particles. 
Since simulations are independent of finite size effects until 
$\Gamma^*(r)$ starts differing from 
$\bar n$ on scales of the order of the system size, the more particles we
have, the larger the system size, the farther in time we can observe the 
evolution of correlations.
For this reason we are now working on a parallelisation of our code, which would 
allow us to simulate systems with more than a million of particles.

\end{document}